\documentclass[aps,prl,reprint,twocolumn,superscriptaddress,showpacs,preprintnumbers,amsmath,amssymb]{revtex4-1}
\usepackage{graphicx}
\usepackage{tabularx}
\usepackage{color}

\usepackage{graphicx}% Include figure files
\usepackage{dcolumn}% Align table columns on decimal point
\usepackage{bm}% bold math

\usepackage[latin1]{inputenc}

\usepackage[T1]{fontenc}
\usepackage{SIunits}
\usepackage{amssymb}
\usepackage{amsfonts}
\usepackage{amsmath}
\usepackage{mathrsfs}
\usepackage{bm}

\begin{document}
\preprint{PREPRINT (\today)}

%opening
\title{Direct observation of the quantum critical point in heavy fermion CeRhSi$_3$}

\author{N.~Egetenmeyer}\email{nikola.egetenmeyer@psi.ch}
\affiliation{Laboratory for Neutron Scattering, Paul Scherrer Institut, CH-5232 Villigen PSI, Switzerland}

\author{J.~L.~Gavilano}
\affiliation{Laboratory for Neutron Scattering, Paul Scherrer Institut, CH-5232 Villigen PSI, Switzerland}

\author{A.~Maisuradze}
\affiliation{Laboratory for Muon Spin Spectroscopy, Paul Scherrer Institut, CH-5232 Villigen PSI, Switzerland}
\affiliation{Physik-Institut der Universität Zürich, Winterthurerstrasse 190, CH-8057 Zürich, Switzerland}

\author{S.~Gerber}
\affiliation{Laboratory for Neutron Scattering, Paul Scherrer Institut, CH-5232 Villigen PSI, Switzerland}

\author{D.~E.~MacLaughlin}
\affiliation{Department of Physics and Astronomy, University of California, Riverside, California 92521, USA}

\author{G.~Seyfarth}
\affiliation{Département de Physique, Univerisité de Montréal, Montréal H3C 3J7, Canada}
\affiliation{Department of Physics and Astronomy, University of California Irvine, Irvine, California 92697-4575, USA}
\affiliation{DPMC, Université de Genève, CH-1211 Geneva, Switzerland}

\author{D.~Andreica}
\affiliation{Faculty of Physics, Babes-Bolyai University, RO-400084 Cluj-Napoca, Romania}

\author{A.~Desilets-Benoit}
\author{A.~D.~Bianchi}
\affiliation{Département de Physique, Univerisité de Montréal, Montréal H3C 3J7, Canada}

\author{Ch.~Baines}
\affiliation{Laboratory for Muon Spin Spectroscopy, Paul Scherrer Institut, CH-5232 Villigen PSI, Switzerland}

\author{R.~Khasanov}
\affiliation{Laboratory for Muon Spin Spectroscopy, Paul Scherrer Institut, CH-5232 Villigen PSI, Switzerland}

\author{Z.~Fisk}
\affiliation{Department of Physics and Astronomy, University of California Irvine, Irvine, California 92697-4575, USA}

\author{M.~Kenzelmann}
\affiliation{Laboratory for Developments and Methods, Paul Scherrer Institut, CH-5232 Villigen PSI, Switzerland}
%N. Egetenmeyer$^{1,2}$, S. Gerber$^{1,2}$, G. Seyfarth$^4$, A. Maisuradze$^3$, D. Andreica$^5$,  A. Desilets-Benoit$^7$, A.D. Bianchi$^7$, D. MacLaughlin$^8$, Ch. Baines$^3$, M. Kenzelmann$^6$, J.L. Gavilano$^1$ \\

\date{\today}% It is always \today, today,
             %  but any date may be explicitly specified

\begin{abstract}
We report on muon spin rotation studies of the noncentrosymmetric heavy fermion antiferromagnet CeRhSi$_3$. A drastic and monotonic suppression of the internal fields, at the lowest measured temperature, was observed upon an increase of external pressure. Our data suggest that the ordered moments are gradually quenched with increasing pressure, in a manner different from the pressure dependence of the Néel temperature. At $\unit{23.6}{kbar}$, the ordered magnetic moments are fully suppressed via a second-order phase transition, and $T_{\rm{N}}$ is zero. Thus, we directly observed the quantum critical point at $\unit{23.6}{kbar}$ hidden inside the superconducting phase of CeRhSi$_3$.
\end{abstract}


\maketitle

\begin{figure}
\centering
\includegraphics[width=\columnwidth]{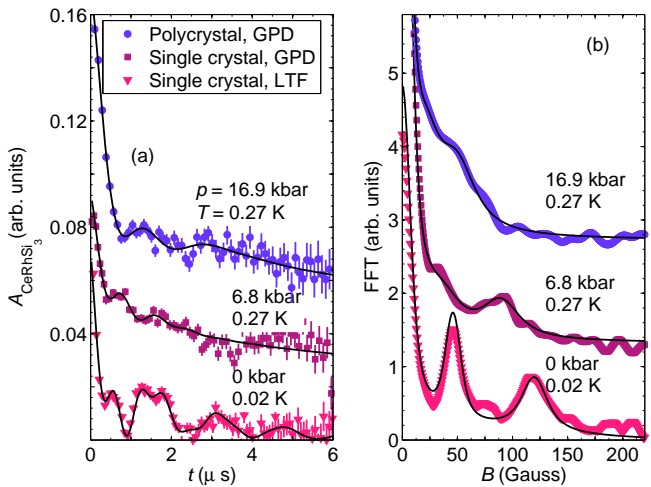}
\caption{ZF measurements at various applied pressures on LTF ($p=\unit{0}{kbar}$) and GPD ($p=\unit{6.8\textrm{ and }16.9}{kbar}$) from single crystal and polycrystalline CeRhSi$_3$ samples. (a) The time spectra at different pressures at base temperature ($\unit{0.02\textrm{ and }0.27}{K}$, respectively) and their corresponding fits (black lines) according to Eq.~(\ref{Eq:LTFequ}). (b) FFTs of the time spectra shown in (a). Both the spectra in (a) and the FFTs in (b) are vertically shifted for better illustration.}\label{GPD1}
\end{figure}

\begin{figure}
\centering
\includegraphics[width=\columnwidth]{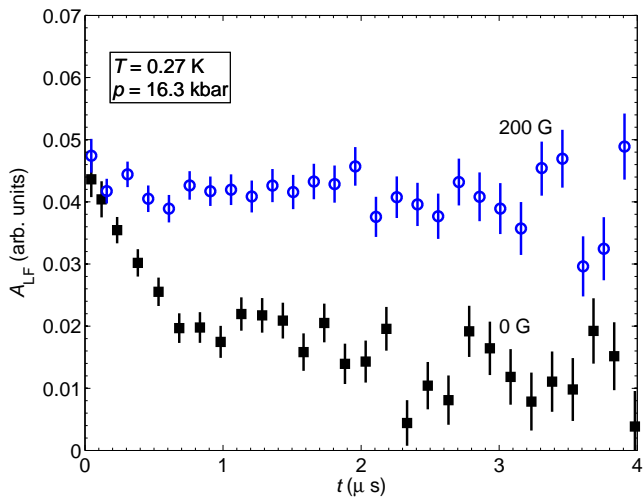}
\caption{$\mu$SR time spectra of the single crystal sample in zero and longitudinal field at $T=\unit{0.27}{K}$ and $p=\unit{16.3}{kbar}$. The pressure cell signal (background) was subtracted.}\label{Fig:LFhP}
\end{figure}

\begin{figure}
\centering
\includegraphics[width=\columnwidth]{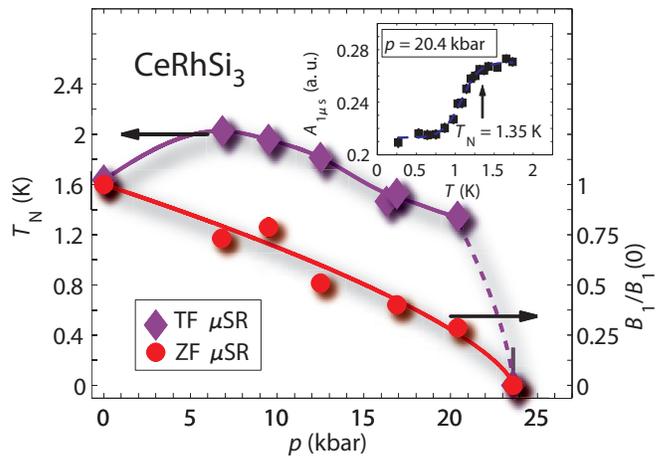}
\caption{Comparison of the pressure dependence of the Néel temperature (diamonds) and the normalized internal field $B_1/B_1(0)$ (circles). The lines are guides to the eye. The inset shows the asymmetry $A_{1 \mu s}$ versus temperature exemplarily for the pressure $p=\unit{20.4}{kbar}$. The Néel temperature is indicated by an arrow. The dashed line is a guide to the eye.}\label{PD}
\end{figure}

%%%%%%%%%%%%%%%%%%%%%%%%%%%%%%%%%%%%%%%%%%%%%%%%%%%%%%%%%%%%%%%%%%%%%%%%%%%%%%%%%%%%%%%%%%%%%%%%%%%%%%%%%%%
\indent The role of magnetism in many superconducting materials is a topic of intense research.
Often, these materials display a quantum critical point (QCP) that separates a nonmagnetic from a magnetic metallic phase~\cite{Gegenwart2008,Si2010,Coleman2005}. This usually leads to the formation of new phases of matter, whose properties are dominated by quantum fluctuations. These fluctuations can be fine-tuned by an external parameter, such as chemical substitution, magnetic field or external pressure. The location and nature of QCPs are central to the understanding of the interplay between superconductivity and magnetism in these systems.
In a number of heavy fermion compounds such as CePd$_2$Si$_2$, CeIn$_3$, and CeIrSi$_3$, the QCP is buried deep inside of a superconducting phase in the pressure-temperature phase diagram~\cite{Si2010,Mathur1998,Settai2008}.
In these cases, it is challenging to establish the presence of a QCP and to study the microscopic magnetic properties coexisting with superconductivity.
In this muon spin rotation ($\mu$SR) study on the heavy fermion CeRhSi$_3$, we uncovered the previously conjectured QCP hidden inside the superconducting phase. We directly observed the disappearance of the ordered magnetic moments at the critical pressure $p^*=\unit{23.6}{kbar}$ deep inside the superconducting phase. This suppression of the ordered moments was found to be accompanied by a zeroing of the Néel temperature at $p^*$ suggesting that this QCP is of magnetic nature.\\
\indent CeRhSi$_3$ is a noncentrosymmetric heavy fermion system~\cite{Kimura,Muro,Muro2007}, suspected to display a QCP~\cite{Kimura,Terashima2007,Tada2010} at pressures of the order of $\unit{26}{kbar}$. The compound features an intriguing phase diagram. It displays an AFM phase at ambient and applied pressure and a superconducting phase for pressures $p>\unit{12}{kbar}$~\cite{Kimura2005,Kimura}.
Measurements of the de Haas--van Alphen effect at ambient and applied pressure show that the \textsl{f} electrons are itinerant up to at least $\unit{30}{kbar}$~\cite{Terashima2007}. Neutron diffraction at ambient pressure showed that the magnetic order is incommensurate, with ordered moments of the order of $\unit{0.1}{\mu_{\rm{B}}/\rm{Ce}}$~\cite{Aso}. From previous results of thermal and transport properties, it was established that the antiferromagnetic ordering temperature $T_{\rm{N}}\approx\unit{1.6}{K}$ at ambient pressure initially increases with increasing pressure and then decreases towards the superconducting transition temperature $T_{\rm{c}}\sim\unit{1}{K}$ for $p\approx\unit{20}{kbar}$. The nature of the magnetism above this pressure across the proposed QCP inside the superconducting phase is unclear.\\
%
%%%%%%%%%%%%%%%%%%%%%%%%%%%%%%%%%%%%%%%%%%%%%%%%%%%%%%%%%%%%%%%%%%%%%%%%%%%%%%%%%%%%%%%%%%%%%%%%%%%%%%%%%%%
\indent Single crystal CeRhSi$_3$ platelets were grown via a standard metal flux technique~\cite{Canfield} with initial molar ratios of 1:1:3:30 (Ce:Rh:Si:Sn).
The high quality of our single crystal samples was verified by the results of x-ray Laue diffraction and by measurements of the residual resistance ratio $\rho_{\unit{300}{K}}/\rho_{0}$, which was found to be of the order of 110.\\
\indent $\mu$SR measurements at ambient pressure were performed by using the Low Temperature Facility spectrometer (LTF) at the Paul Scherrer Institut.
Measurements under pressure were conducted on the general purpose decay-channel spectrometer (GPD).
For zero-field (ZF) and longitudinal-field (LF) measurements, we typically counted $6\times10^6$ positron events per spectrum and $0.5\times10^6$ for transverse-field (TF) measurements, respectively.\\
\indent The single crystal sample consisted of platelets with an average thickness of $\unit{0.1}{mm}$. For the measurement on LTF, they were attached to a silver plate, with the crystal \emph{c} axis perpendicular to the plate and parallel to the muon beam covering an area of $\sim \unit{1}{cm^2}$.
For pressure measurements, they were stacked in a cylindrical volume of $\unit{1}{cm^3}$ (mass $\sim\unit{500}{mg}$).
In addition, a $\unit{1}{g}$ polycrystalline sample of CeRhSi$_3$ was measured under pressure.\\
%
%%%%%%%%%%%%%%%%%%%%%%%%%%%%%%%%%%%%%%%%%%%%%%%%%%%%%%%%%%%%%%%%%%%%%%%%%%%%%%%%%%%%%%%%%%%%%%%%%%%%%%%%%%%
\indent In Fig.~\ref{GPD1}, ZF time spectra and the corresponding fast Fourier transforms (FFTs) for several pressures at a base temperature (of $\unit{0.02}{}$ and $\unit{0.27}{K}$ for LTF and GPD, respectively) are shown for single crystal and polycrystalline samples. The signal from the Ag plate and the pressure cell, respectively, has been subtracted (see Supplemental Material for details).
The signals shown in Fig.~\ref{GPD1} reveal a fast decay and a characteristic time structure, expected in the case of a signal containing several oscillatory components. The FFT of the time spectrum at $p=\unit{0}{kbar}$ (see Fig.~\ref{GPD1}(b)) displays three well-defined distributions of internal magnetic fields centered at $\sim\unit{0 \textrm{, }45 \textrm{, and }120}{G}$.
%The observed $\mu$SR signals are composed of the signal from the sample and an additional background signal.
Therefore, the ZF sample asymmetry function is described here by three terms:
\begin{eqnarray}\label{Eq:LTFequ}
&&A_{\rm{CeRhSi_3}}(t)=\sum_{i=1}^{3}A_{i}\cdot g_i(t),\\ \nonumber
&&g_i(t)=\alpha\cdot\cos{(\gamma_{\mu}B_i t)}\cdot e^{-\lambda_{\rm{T}i}\cdot t}
+(1-\alpha)\cdot e^{-\lambda_{\rm{L}}\cdot t}, \nonumber
\end{eqnarray}
where $\gamma_{\mu}/2\pi=\unit{135.5}{MHz/T}$ with $\gamma_{\mu}$ the muon gyromagnetic ratio. $B_i$ denote the average internal magnetic fields at the muon stopping sites in the sample. $\alpha$ was found to be $\frac{2}{3}$, which is not \emph{a priori} evident for single crystals but is always the case for polycrystalline samples. $A_{i}$ are the asymmetries at time 0 of the individual components, which are proportional to the probability that a muon stops at a site with an internal field $B_i$. We find $A_1=A_2$ and $A_3=1.5\cdot A_1$. $\lambda_{\rm{T}i}$ and $\lambda_{\rm{L}}$ are the transverse and longitudinal muon relaxation rates, respectively. $B_i$, $\lambda_{\rm{T}i}$ and $\lambda_{\rm{L}}$ are temperature-dependent. The relations $B_1/B_2 \approx 2.6$ and $\lambda_{\rm{T}1}/\lambda_{\rm{T}2} \approx 2.6$ were found experimentally; $B_3$ was found to be roughly $0$ for all temperatures and pressures.\\
\indent The weights of the spectra in Fig.~\ref{GPD1}(b) shift to lower fields with increasing pressure; that is, the quasistatic internal fields are increasingly suppressed with pressure.\\
\indent Since the AFM structure is incommensurate, one might expect that the quality of the fits would improve by replacing the cosine terms in Eq.~(\ref{Eq:LTFequ}) by Bessel functions, as in the case of UNi$_2$Al$_3$~\cite{UemuraLuke,AmatoRev}, CeAl$_3$~\cite{AmatoRev} or CeCu$_5$Au~\cite{Amato1995}. This replacement, however, leads to fits of lower quality and will not be considered further here.\\
%
%%%%%%%%%%%%%%%%%%%%%%%%%%%%%%%%%%%%%%%%%%%%%%%%%%%%%%%%%%%%%%%%%%%%%%%%%%%%%%%%%%%%%%%%%%%%%%%%%%%%%%%%%%%
\begin{figure}
\centering
\includegraphics[width=\columnwidth]{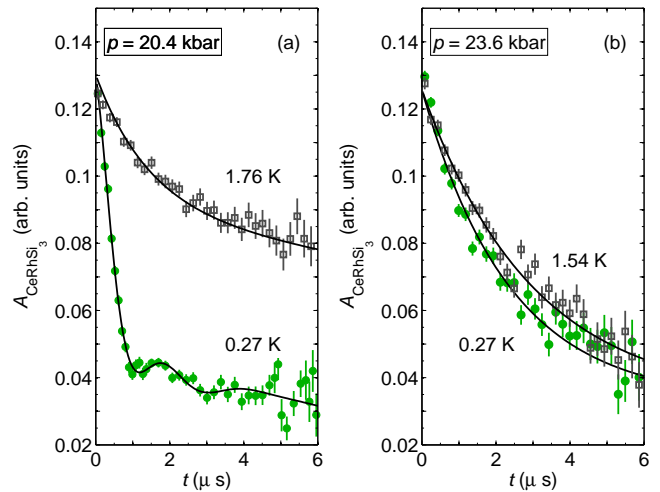}
\caption{ZF data from the polycrystalline sample at (a) $\unit{20.4}{}$ and (b) $\unit{23.6}{kbar}$ at two different temperatures. The time spectra were fitted by using Eq.~(\ref{Eq:LTFequ}). The signal of the pressure cell was subtracted.}\label{Fig:HighPressure}
\end{figure}
\indent High-pressure data at low and high temperatures are shown in Fig.~\ref{Fig:HighPressure}.
In the time spectrum at $\unit{20.4}{kbar}$ and $T=\unit{0.27}{K}\ll T_{\rm{N}}$ we observe a fast relaxation and slow oscillations. Both of these features are absent at $T=\unit{1.76}{K}> T_{\rm{N}}$ showing that the low-temperature data reflects the presence of internal magnetic fields arising from magnetic order. For the data at $\unit{23.6}{kbar}$, the fast relaxation and the oscillations are absent, and there is no significant change from the high temperature to the low-temperature signal at $T=\unit{0.27}{K}$.
%(The fast relaxation in the normal state at $\unit{23.6}{kbar}$ which is much less pronounced in the data at $\unit{20.4}{kbar}$, we believe, is associated with an imperfect subtraction of the MP35 pressure cell for the highest pressure. This is a known difficulty of this technique.)
This demonstrates the absence of magnetic order down to at least $\unit{0.27}{K}$, the lowest temperature of these experiments. Since at this pressure the internal magnetic fields are zero, $T_{\rm{N}}$ is most likely zero.\\
\indent In the results of resistivity and specific heat measurements, CeRhSi$_3$ was found to be superconducting for $p>\unit{12}{kbar}$ and low $T$; our results therefore demonstrate that over the pressure region $\unit{12}{kbar}<p<\unit{23.6}{kbar}$ superconductivity and AFM order coexist, as indicated in earlier measurements~\cite{Kimura}.\\
%%%%%%%%%%%%%%%%%%%%%%%%%%%%%%%%%%%%%%%%%%%%%%%%%%%%%%%%%%%%%%%%%%%%%%%%%%%%%%%%%%%%%%%%%%%%%%%%%%%%%%%%%%%
\indent The ZF data in the AFM state show, in addition to the tiny ordered moments, a fast initial relaxation. As we shall see, the fast relaxation is not caused by fluctuating (large) moments, as it would be in the case where Kondo screening was destroyed and the local Ce moments were restored. The results of $\mu$SR measurements at longitudinal fields at $p=\unit{16.3}{kbar}$ (see Fig.~\ref{Fig:LFhP}) showed that at modest external fields of $\unit{200}{G}$ (of the order of the internal quasistatic fields) the muon relaxation is substantially suppressed. That means that the $\mu^{+}$ spin is fully decoupled from the internal magnetic field, demonstrating that the weak internal fields are static rather than dynamic and supporting the quasistatic origin of the relaxation at the pressure $\unit{16.3}{kbar}$.
%that is, magnetism is fully decoupled, which supports the quasi static origin of the relaxation at the pressure $\unit{16.3}{kbar}$.
We find no evidence for dynamical relaxation at any pressure in the low-temperature region of the AFM phase. Above $T_{\rm{N}}$ at $p=\unit{0}{kbar}$ an LF measurement was also conducted and complete decoupling from internal fields was observed at an applied field of $\sim\unit{200}{G}$, showing the absence of appreciable dynamical relaxation above $T_{\rm{N}}$. %We therefore infer absence of appreciable dynamical relaxation above $T_{\rm{N}}$.
Thus, our data support a scenario where the Ce moments are quenched at all pressures below $p^*$ and where the Kondo screening remains largely intact.\\
%%%%%%%%%%%%%%%%%%%%%%%%%%%%%%%%%%%%%%%%%%%%%%%%%%%%%%%%%%%%%%%%%%%%%%%%%%%%%%%%%%%%%%%%%%%%%%%%%%%%%%%%%%%
\indent The normalized static internal fields $B_1/B_1(0)(p)$ [equivalent to $B_2/B_2(0)(p)$] extracted from the fits using Eq.~(\ref{Eq:LTFequ}) are depicted in Fig.~\ref{PD} as a function of pressure (circles).
$B_1(0)$ denotes the internal field at the lowest measured pressure $p=\unit{0}{kbar}$.
$B_{1}$ decreases monotonically with pressure and vanishes near $p^{*}\sim\unit{23.6}{kbar}$, showing that magnetic order is suppressed at $p^*$. For the two highest measured pressures, the field distribution is very broad. Still, over the entire antiferromagnetically ordered region, one observes a pressure-independent ratio $B_1/B_2$. This implies that the magnetic structure does not change appreciably with pressure. In addition, our data show that near $p^*$ the fast relaxation characterizing the ZF data in the AFM state vanishes.\\
%%%%%%%%%%%%%%%%%%%%%%%%%%%%%%%%%%%%%%%%%%%%%%%%%%%%%%%%%%%%%%%%%%%%%%%%%%%%%%%%%%%%%%%%%%%%%%%%%%%%%%%%%%%
\indent The Néel temperature as a function of pressure was obtained with $\mu$SR measurements in transverse fields of $\unit{50 \textrm{ and } 300}{G}$, respectively. Below the Néel temperature the magnetic order leads to an extra fast relaxation, so that the total asymmetry after some time, $t>1\mu s$ in our case, is smaller than the corresponding amplitude above $T_{\rm{N}}$. The time spectra for $t>\unit{1}{\mu s}$ were fitted by using
\begin{equation}
A(t)=A_{1 \mu s} \cdot \exp{\big(-\frac{1}{2}(\sigma t)^2\big)}\cdot\cos{(\gamma_{\mu}B t)},
\end{equation}
where $A_{1\mu s}$ is the asymmetry at 1 $\mu s$ after the muon is implanted. $B$ represents the local field and is a fit parameter. Upon cooling, one observes a rapid decrease of $A_{1 \mu s}$, as the system enters the AFM state.
The Néel temperature is associated with the onset of the drastic steplike reduction of $A_{1\mu s}(T)$ with decreasing temperature. One example of $A_{1 \mu s}(T)$ for $p = \unit{20.4}{kbar}$ is shown in the inset in Fig.~\ref{PD}.\\
\indent The results for $T_{\rm{N}}$ at different pressures are shown in Fig.~\ref{PD} as diamonds. $T_{\rm{N}}(p)$ shows a maximum at around $\unit{7}{kbar}$, consistent with the results of resistivity~\cite{Kimura2005,Kimura} and specific heat~\cite{Tomioka2007} measurements. By contrast, the internal fields and, therefore, the ordered magnetic moments decrease monotonically.
We find that the internal fields and $T_{\rm{N}}$ vanish at the critical pressure of $p^{*}=\unit{23.6}{kbar}$ and magnetic order disappears.
Our data show that the AFM order is completely suppressed for $p\geq p^{*}$ and that there is a continuous tuning of the AFM phase with pressure. All this indicates the presence of a magnetic QCP at $p^{*}$.\\
\indent The different pressure dependencies of the Néel temperature and that of the internal fields may be understood by using the simplified, but widely used, one-dimensional Kondo lattice model described by Doniach~\cite{Doniach}. The model considers the competing effects of the Kondo screening and the exchange interaction between Ce moments which tend to stabilize the magnetic moments leading to magnetic order. The results are summarized in the so-called Doniach phase diagram, displaying the Néel temperature $T_{\rm{N}}$ as a function of the product $|JN_{\rm{F}}|$ of the exchange coupling $J$ and the density of states of the conduction electrons at the Fermi level $N_{\rm{F}}$.
For low values of $|JN_{\rm{F}}|$, the system is magnetically ordered and $T_{\rm{N}}$ increases with increasing $|JN_{\rm{F}}|$. Further increase of $|JN_{\rm{F}}|$ yields a maximum of $T_{\rm{N}}$ and a subsequent decrease (observed, for instance, in CeAg~\cite{Eiling}). $T_{\rm{N}}$ eventually reaches zero, and a further increase in $|JN_{\rm{F}}|$ yields a nonmagnetic (Fermi liquid) system.\\
\indent In our case, the product $|JN_{\rm{F}}|$ is tuned with external pressure. The observation of the maximum of $T_{\rm{N}}(p)$ and the subsequent suppression of $T_{\rm{N}}$ to zero suggest that the application of rather modest pressures leads to substantial changes of $|JN_{\rm{F}}|$. At the same time, the Doniach model predicts a monotonic decrease of the ordered moments with increasing $J$ according to~\cite{Doniach} $\langle S^x\rangle=\sqrt{1-(J/W)^2}$, where $W$ is the conduction electron bandwidth and $S^x$ is the $x$ component of a spin $\bm{S}$. These predictions match qualitatively our experimental observations.\\
%\indent The critical pressure $p^{*}$ marks the transition point from the magnetically ordered phase to a nonmagnetic state.
\indent We emphasize that the order parameter, i.e., the sublattice magnetization (or more precisely the amplitude of the spin density wave), represented here by the internal fields, displays a continuous phase transition with pressure and the Néel temperature goes to zero at $p^*$. This clearly reveals the presence of a QCP. In our case, it lies deep in the superconducting phase, and its bulk properties are masked by superconductivity. Nevertheless, it has been reported that between $\unit{20}{}$ and ${26}{kbar}$ and $T>\unit{1}{K}$, non-Fermi-liquid behavior~\cite{Kimura} is observed in the temperature dependence of the electrical resistivity, adding support to our claim. Further indications for a QCP may be extracted from the extremely high upper critical field $H_{\rm{c}2}$ ($H_{\rm{c}2}\|\bf{c}$) and its sensitivity to pressure for $p  > \unit{26}{kbar}$~\cite{Kimura2007}.
%, a downward curvature in the temperature dependence $H_{\rm{c}2}(T)$, and the strong changes of $H_{\rm{c}2}$ with pressure.
Both the behavior of the resistivity and that of $H_{\rm{c}2}$ may be explained theoretically in terms of antiferromagnetic spin fluctuations in three dimensions close to the QCP~\cite{Tada2010}.
%-------add for 2nd referee
From the results of transport measurements, it was not clear what the critical pressure should be. Nevertheless, the authors suggest that $p^{*}$ is about $\unit{26}{kbar}$~\cite{Kimura2007}, a value not compatible with our findings.
%---------------------------
In the results of de Haas--van Alphen measurements, no change in the size of the Fermi surface across $p^*$ was observed \cite{Terashima2007}. That is, the Kondo screening of the Ce moments is active at pressures on both sides of $p^*$. All this hints at a spin-density-wave--type QCP~\cite{Si2010,Gegenwart2008}, consistent with our findings. This type of QCP is driven by quantum fluctuations of the magnetic order parameter, as opposed to the so-called local QCP, where the Kondo screening is destroyed and the local moments are restored in the magnetically ordered phase, i.e., $p<p^{*}$.\\%, not involving a breakdown of the Kondo shielding across the QCP.\\
%
%In our own $\mu$SR measurements the magnetic moments were found to be rather small ($<0.3\mu_{\rm{B}}$). Additionally in the paramagnetic state the tiny relaxation in the signal could be decoupled as observed in LF measurements. Thus there is no indication for local moments on the AFM side of the QCP. Therefore we conclude that Kondo screening is still present even for pressures $p\leq p^{*}$. The QCP in CeRhSi$_3$ is of the SDW type, not involving a Kondo breakdown.\\
%
%%%%%%%%%%%%%%%%%%%%%%%%%%%%%%%%%%%%%%%%%%%%%%%%%%%%%%%%%%%%%%%%%%%%%%%%%%%%%%%%%%%%%%%%%%%%%%%%%%%%%%%%%%%
\indent In summary, we have measured the pressure evolution of the internal fields of the AFM phase of CeRhSi$_3$ with the $\mu$SR technique.
As a function of pressure the internal fields diminish continuously with increasing pressure, in clear contrast to the complex behavior of the Néel temperature, which shows a maximum at around $\unit{7}{kbar}$. Our observations may be understood in the framework of the Doniach phase diagram. This analysis reveals a strong sensitivity of $|JN_{\rm{F}}|$ to the externally applied pressure for CeRhSi$_3$. At $p^{*}\approx\unit{23.6}{kbar}$, the system becomes nonmagnetic and signals the presence of a magnetic QCP, masked by the superconducting phase. We find no evidence for a breakdown of the Kondo screening across the QCP.\\
%
%%%%%%%%%%%%%%%%%%%%%%%%%%%%%%%%%%%%%%%%%%%%%%%%%%%%%%%%%%%%%%%%%%%%%%%%%%%%%%%%%%%%%%%%%%%%%%%%%%%%%%%%%%%
\indent This work was performed at the Swiss Muon Source (S$\mu$S), Paul Scherrer Institut (PSI Villigen, Switzerland). We thank A. Amato and H. Luetkens for helpful discussions and M. Elender for technical support. We acknowledge the financial support of the Swiss National Science Foundation. D.~A. acknowledges financial support from the Romanian UEFISCDI Project No. PN-II-ID-PCE-2011-3-0583 (85/2011). A.~D.~B. received support from the Natural Sciences and Engineering Research Council of Canada (Canada), Fonds Québécois de la Recherche sur la Nature et les Technologies (Québec), and the Canada Research Chair Foundation. Work at UC Riverside was supported by U.S. NSF Grant No. 0801407.

%%%%%%%%%%%%%%%%%%%%%%%%%%%%%%%%%%%%%%%%%Supplemental Material%%%%%%%%%%%%%%%%%%%%%%%%%%%%%%%%%%%%%%%%%%%%%%
\appendix\section*{Supplemental material: Background signal}
%\appendix\section*{Signal from the pressure cell}\label{MyAppendix}
\indent In our experiments the raw $\mu$SR signal contains two components: the signal from the sample, CeRhSi$_3$, and the signal from the background (bg), i.e.
\begin{equation}\label{Eq:general}
A(t)=A_{\rm{CeRhSi_3}}(t)+A_{\textrm{bg}}(t).
\end{equation}
In LTF the background signal originates from muons stopped in the silver plate to which the crystals are attached, whereas in GPD the background stems from muons stopped in the pressure cell. Both cases have been extensively characterized in previous experiments and their relaxation is due to the nuclear moments of the materials. In the analysis of our data, the relaxation rates of the respective backgrounds were fixed in the fits and the only free parameter is the initial asymmetry $A_{\rm{bg}}(0)$. With this restriction we perform a full fit to the data using Eq.~\ref{Eq:general}, from where we extract $A_{\rm{CeRhSi_3}}$.\\
\indent The silver plate used in the LTF experiment contributes a signal that is roughly constant in the measured time window. Approximately $50\%$ of muons stopped inside of the sample in the LTF measurements.\\
\indent For the GPD measurements, we used piston-cylinder pressure cells made of CuBe ($p\leq \unit{21}{kbar}$) and the Co-Ni alloy MP35 ($p>\unit{21}{kbar}$), respectively~\cite{Andreica}. The hydrostatic pressure was measured with an $\unit{0.1}{kbar}$ accuracy in the following way: with ac-susceptibility measurements the superconducting transition temperature $T_{\rm{c}}$ of an indium piece, placed inside the pressure cell, was determined. The pressure dependence of the superconducting transition of In is
\begin{equation}
T_{\rm{c}}=\unit{3.402}{K}-\unit{3.64\cdot10^{-2}}{K/kbar}\cdot p,
\end{equation}
where $p$ is the pressure in kbars~\cite{Maisuradze2011}.\\
\indent The pressure cell contributes a rather high background compared to the signal from the sample. For the single crystal sample only 20 - 25\% of the incoming muons stopped in the sample. For the polycrystalline sample the stopping fraction increased to 40\%.\\
\indent In zero field (ZF) the signal due to the CuBe-pressure cell can be described by a Kubo-Toyabe type polarization function~\cite{Kubo}:
%[Eq.~8 of \onlinecite{Hayano}]
\begin{equation}\label{KT}
A_{\rm{KT}}(t)=\frac{1}{3}+\frac{2}{3}(1-(\sigma t)^2)\exp{[-\frac{1}{2}(\sigma t)^2]},
\end{equation}
where $\sigma/\gamma_{\mu}$ is the width of the local field distribution with $\sigma=\unit{0.33}{\mu s^{-1}}$ (temperature independent) for CuBe.\\
% The muon depolarization is due to Be and Cu nuclear moments~\cite{Andreica}.\\
%of Cu$^{63}$, Cu$^{65}$ and Be$^9$
\indent For the MP35-pressure cell the signal in ZF may be approximated by the product of a Kubo-Toyabe (Eq.~(\ref{KT})) and an exponential function, i.e.~\cite{Andreica,Khasanov2009}
\begin{equation}\label{Eq:MP35}
A_{\textrm{MP35(ZF)}}(t)=A_{\rm{KT}}(t)\cdot\exp{(-\lambda t)}.
\end{equation}
The relaxation rate $\lambda$ is found to be temperature dependent and was fixed at each temperature to values from a calibration curve. The calibration curve was obtained at the beginning of the experiment at the measured pressure. For this purpose the energy of the incoming muons is adapted so that the muons stop only inside of the pressure cell (and not in the sample). In a weak transverse field (TF) the pressure cell is fitted by
\begin{equation}
A_{\textrm{MP35(TF)}}=A_0 \exp[{-\frac{1}{2}(\sigma' t)^2}] \exp({-\lambda t}) \cos{(\gamma_{\mu}B t)},
\end{equation}
with $A_0$ the asymmetry at time zero, $\sigma'\approx\unit{0.33}{\mu s^{-1}}$ (temperature independent) and $B$ the local field parameter. $\lambda(T)$ is marked by a strong increase below $T =\unit{1.4}{K}$. The subtraction of the MP35-pressure cell signal in the low temperature region therefore requires special care, due to its strong temperature dependence.\\
\indent The signal of the (CuBe) pressure cell in a longitudinal field (LF) is described by an extended Kubo-Toyabe type function [Eq.~21 of Ref.~\onlinecite{Kubo} and Eq.~6.130 of Ref.~\onlinecite{Yaouanc}]:
%[Eq.~21 of Ref.~\onlinecite{Kubo} and Eq.~2 of Ref.~\onlinecite{Alex}] :
\begin{eqnarray}\label{Eq:PC_LF}
A_{\rm{LF}}(t)&=&1-\frac{2Q'(t)}{\omega^2t}[\cos{\omega t}-j_0(\omega t)]\nonumber\\
&-&2\int_0^t~\frac{Q'(s)}{s} \frac{j_0'(\omega s)}{\omega}~\rm{d}s,
\end{eqnarray}
with $j_0(x)=\sin(x)/x$ and $Q(t)=\exp(-\frac{1}{2}\sigma^2t^2)$.\\

%------------------------------------ REFRERENCES
%\bibliographystyle{apsrev4-1}
%\bibliography{CeRhSi3bib_etal}

%merlin.mbs apsrev4-1.bst 2010-07-25 4.21a (PWD, AO, DPC) hacked
%Control: key (0)
%Control: author (72) initials jnrlst
%Control: editor formatted (1) identically to author
%Control: production of article title (-1) disabled
%Control: page (0) single
%Control: year (1) truncated
%Control: production of eprint (0) enabled
%

\end{document}